\documentclass[sigconf,authorversion,nonacm]{acmart}

\AtBeginDocument{%
  \providecommand\BibTeX{{%
    \normalfont B\kern-0.5em{\scshape i\kern-0.25em b}\kern-0.8em\TeX}}}

\usepackage[printonlyused,nohyperlinks]{acronym}
\usepackage{amsmath,amsfonts}

\usepackage{graphicx}
\usepackage{algorithm}
\usepackage{textcomp}
\usepackage{xcolor}

\usepackage{glossaries}

\usepackage{pifont}
\usepackage{graphicx}
\usepackage{textcomp}
\usepackage{xcolor}
\usepackage{amsmath}
\usepackage{amsthm}
\usepackage{ gensymb }
\usepackage{ upgreek }
\usepackage{xspace}
\usepackage{longtable}
\theoremstyle{plain}
\usepackage{array}

	\acrodef{CPS}{Cyber Physical Systems}
	\acrodef{ACAS-Xu}{Collision Avoidance System-Xu}
	\acrodef{APS}{Artifical Pancreas System}
	\acrodef{AV}{Autonomous Vehicles}
	\acrodef{CA}{Collision Avoidance}
	\acrodef{CPS}{Cyber Physical Systems}
	\acrodef{CGM}{Continuous Glucose Monitor}
	\acrodef{DNN}{Deep Neural Network}
	\acrodef{FDIA}{False Data Injection Attacks}
	\acrodef{ReLUSyn}{ReLU Synthesizer}
	\acrodef{HCAS}{Horizontal Collision Avoidance System}
	\acrodef{LP}{Linear Programs}
	\acrodef{MILP}{Mixed Integer Linear Program}
	\acrodef{MIP}{Mixed Integer Programs}
	\acrodef{ML}{Machine Learning}
	\acrodef{NN}{Neural Network}
	\acrodef{RFDIA}{Ripple False Data Injection Attacks}
	\acrodef{RPi}{Raspberry Pi}
	\acrodef{ReLU}{Rectified Linear Unit}
	\acrodef{SMT}{Satisfiability Modulo Theories}

\setcopyright{none}
\begin{document}

\title{ReLUSyn: Synthesizing Stealthy Attacks for Deep Neural Network Based Cyber-Physical Systems}

\author{Aarti Kashyap}
\authornote{Both authors contributed equally to this research.}
\email{aartisk@cs.ubc.ca}
\author{Syed Mubashir Iqbal}
\authornotemark[1]
\email{syedmubashiriqbal@gmail.com}
\affiliation{%
	\institution{University of British Columbia, Vancouver}
}

\author{Karthik Pattabiraman}
\affiliation{%
	\institution{University of British Columbia}}
\email{karthikp@ece.ubc.ca}

\author{Margo Seltzer}
\affiliation{%
	\institution{University of British Columbia}}
\email{mseltzer@cs.ubc.ca}

\begin{abstract}
\ac{CPS} are deployed in many mission-critical settings, such as medical devices, autonomous vehicular systems and aircraft control management systems.
	As more and more CPS adopt Deep Neural Networks (\ac{DNN}s),
these systems can be vulnerable to attacks.
.
Prior work has demonstrated the susceptibility of CPS to False Data Injection Attacks (\ac{FDIA}s),
which can cause significant damage.
We identify a new category of attacks on these systems.
In this paper, we demonstrate that DNN based CPS are also subject to these attacks.
These attacks, which we call \ac{RFDIA}, use minimal input perturbations to
stealthily change the \ac{DNN} output.
The input perturbations propagate as ripples through multiple \ac{DNN} layers
 to affect the output in a targeted manner. 
 
We develop an automated technique to synthesize \ac{RFDIA}s against DNN-based CPS.
	Our technique models the attack as an optimization problem using Mixed Integer Linear Programming (\ac{MILP}).
We define an abstraction for \ac{DNN}-based \ac{CPS} that allows us to automatically: 
	1) identify the critical inputs, and 2) find the smallest perturbations that produce output changes. 
We demonstrate our technique on three practical \ac{CPS} with two mission-critical applications: an (\ac{APS}) and two aircraft control management systems (\ac{HCAS} and \ac{ACAS-Xu}). 

\end{abstract}




\maketitle

\newcommand{\tool}{{\em ReLUSyn }}
\newcommand{\attack}{{\em ripple FDI attack }}

\section{Introduction }
\label{ch:Chapter1}

CPS are increasingly deployed for 
safety-critical applications \cite{10.1145/2038642.2038685}\cite{10.1145/1837274.1837463}
\cite{6051465} 
such as medical devices (e.g., insulin pumps and robotic surgery equipment), for which failures are life threatening and hence need high reliability. 
The mission-critical aspect of \ac{CPS} makes them particularly attractive targets for attackers. 
There have been multiple attacks demonstrated on \ac{CPS} in the past.
e.g., targeted attacks on surgical robots \cite{7579758}, radio attacks on pacemakers\cite{4531149}, airplanes \cite{217595} and  cars \cite{10.5555/1929820.1929848}.

Traditionally, \ac{CPS} have used classical control theory-based models \cite{6051465} \cite{1337806} \cite{10.1145/2038642.2038667}  to calculate control actions. 
These systems model the physical world via differential equations with 

parameters (or variables) such as road friction, wind speed and direction, and engine power. 
Such models have low tolerance for perturbations in the values of these parameters, because the equations account for a subset of all possible behaviors that can happen in reality. 
This behavior makes \ac{CPS} vulnerable to attacks that can inject a false reading of a variable \cite{10.1145/1952982.1952995}. 
Thus, an adversary can cause irreparable harm to such systems by injecting a false variable value; 
Such attacks are called \ac{FDIA}.
In an \ac{FDIA}, an attacker compromises the outputs from the sensors to
produce errors in the output that go undetected \cite{7438916}.

Recently,  \ac{DNN} based controllers have replaced classical control-theory based controllers \cite{xiang18} \cite{Kocic2019} \cite{bechtel2017deeppicar}.  
Techniques used to generate inputs to mount an FDIA attack for classical control theory based controllers are not directly applicable for DNNs, because the state space for DNN based systems is typically much larger than that for classical control theory models. 

Further, in a \ac{DNN}, the computation proceeds through multiple layers making the input-output mapping opaque.

We demonstrate in this paper that even in the presence of a \ac{DNN},
a motivated attacker can launch \ac{FDIA}s.

This paper introduces an {\em automated technique} to synthesize the attacks for \ac{DNN} based controllers that perturb specific inputs to produce changes in output. 
We call these attacks Ripple FDIAs (\ac{RFDIA}s), as they are transmitted through the DNN's layers in the CPS much like ripples in water puddles.

To mount a successful \ac{RFDIA} attack on a DNN based controller, the attacker needs to minimally distort an input such that the effects of its perturbation reach the final layer,
without triggering any alarms. This involves identifying 
(1)	\textit{The critical inputs}: The inputs (for the attacker) that can be perturbed (either together or in isolation) to cause a desirable change in the output, and 
(2) \textit{Perturbations for each critical input}: The smallest perturbations that will trigger the desired change in output.

We present an automated technique that finds the \textit{critical inputs} and the \textit{minimal perturbations} to cause a successful \ac{RFDIA}. 
We model attack synthesis as an optimization problem using Mixed Integer Linear Programming (\ac{MILP}), and build an automated tool \tool that identifies the critical inputs and finds the perturbations.
{\em To the best of our knowledge, we are the first to  propose an automated approach to systematically identify the critical inputs and find the minimum perturbations to conduct an \ac{RFDIA} for DNN-based controllers.}

The main contributions in this work are as follows:

\begin{enumerate}
	\item Define and demonstrate a new class of attacks called \ac{RFDIA} for \ac{DNN} based \ac{CPS}. 
	\item Model \ac{RFDIA} synthesis using \ac{MILP}, and implement it as a tool, \tool. 
	\item Demonstrate \tool on three \ac{DNN} based \ac{CPS}:  An \ac{APS} and two air traffic control management systems called \ac{ACAS-Xu} and \ac{HCAS}.
	\item Evaluate \tool's performance to quickly identify the critical inputs and
	perturbations for mounting an \ac{RFDIA}.
\end{enumerate}

The main results in this work is that \tool exposed  many vulnerabilities in \ac{DNN} based systems. 
It found approximately 7000, 20, and 40 successful
\ac{RFDIA}s for \ac{APS}, \ac{HCAS} and \ac{ACAS-Xu} in fewer than
1 second, 40 seconds, and 7000 seconds, respectively. Our work thus points the way to make the networks more resilient to \ac{RFDIA}s.

\section{Related Work}
\label{relatedwork}

We have classified related work into four broad categories. 
The first category discusses the verification techniques used
to verify DNNs, highlighting
why these techniques cannot directly be used for our application.
The second subsection discusses work from the \ac{FDIA} community for
classic control theory modeled systems, illustrating why such approaches
do not apply when \ac{DNN}s replace class control theory.
In the third subsection, we discuss applications where DNNs have replaced
or have been added to the CPS.
This provides a glimpse into how the DNNs are being used in safety-critical
applications, further motivating the need for a technique that can generalize
to different DNN-based CPS.
Finally, in the last subsection, we discuss adversarial attacks that have
attracted attention in the Machine Learning (ML) community and how these
differ from our \attack.

\subsection{Verification of CPS}

Formally verifying DNN-based CPS has recently gained a lot of momentum. Researchers have proposed automated verification mechanisms that use techniques involving SMT solvers, symbolic execution and MILP \cite{10.1007/978-3-642-14295-6_24}, \cite{article}, \cite{10.1007/978-3-319-63387-9_5}. 
The goal of these techniques is to conduct an input-output range analysis to establish bounds for the DNN's correctness. Our work focuses not on conducting a bound analysis, but on finding the inputs for changing the outcome of a CPS under a constrained setting for a FDIA.
Additionally, existing approaches focus more on verifying certain properties
such as reachability analysis that do not account for \attack.

Pulina et al.\cite{10.1007/978-3-642-14295-6_24} proposed one of the first approaches for verifying DNN safety properties using abstraction refinement.
They abstract \cite{article} the nonlinear activation functions in a linear arithmetic \ac{SMT} formula, but their application and abstraction interface is different from ours. 
They then use a counterexample-based approach for abstract refinement.
There has been further work by Katz et al.\cite{10.1007/978-3-319-63387-9_5} in verifying DNNs using SMT solvers where they build rules for handling the ReLU activation function.
Much of the work in this domain utilizes the state-of-the art solvers Z3 and CVC4 to prove robustness properties \cite{NIPS2016_6339} for DNNs, but cannot detect or produce FDIA.
Xiang at al.\cite{xiang2017output} focus on an output reachable estimation for NNs using simulation-based methods.
There have been multiple methods proposed to conduct reachability analysis for DNNs using a MILP/SMT approach \cite{10.1145/3302504.3313351} \cite{ehlers2017formal} \cite{10.1007/978-3-319-63387-9_5} \cite{lomuscio2017approach} \cite{article}.
Ivanov et al. \cite{ivanov2018verisig} propose a verification framework for
verifying DNN based controllers by abstracting the problem into a hybrid system verification problem.

\subsection{False Data Injection Attacks (FDIA) in traditional systems}
Attacking a CPS by exploiting the deep integration of the physical-nature and cyber-nature of the systems has lead to the emergence of FDIAs. 
There are two main categories of FDI attacks that have been proposed in the literature: random FDIA and targeted FDIA. Random FDIA are generated by randomly changing the inputs to cause changes in the state estimation.
Targeted FDIAs are more challenging from an attacker's perspective, because they must synthesize exact values for inputs to cause a change in the output without triggering alarms.
We also synthesize both random and targeted FDIAs using
constraints from the specification documentation. 

In FDI attacks, the attacker's goal is to compromise the sensor inputs to mislead the state estimation process \cite{e3f0020abba24d4389aff937fe8bcdd5}. Liu et al. \cite{10.1145/1952982.1952995} introduced the class of FDIA for electric power grids to introduce arbitrary errors into certain state variables without being detected. Giannini et al. \cite{unknown} generate FDIA by introducing new information as constraints in the state estimator under the assumption that the new information is not available to the attacker. These techniques generate FDIA for conventional control systems, causing
wrong state estimation, while avoiding detection.
Our work addresses a similar problem for \ac{DNN}-based CPS.
Bobba et al.
strategically selected set of sensor measurements, it is possible to detect
the attacks proposed by Liu at al. \cite{10.1145/1952982.1952995}.

\subsection{DNN controllers in CPS}
This line of work focuses on intelligent control using \ac{DNN}.
Fukuda et al. \cite{170966} show how NNs can be utilized in industrial systems. Cong et al. \cite{Cong} propose a recurrent neural network to create a Proportional–Integral–Derivative (PID, K$_p$, K$_i$ and K$_d$)  neural network (PIDNN). 
This allows the network to converge more quickly since PIDNN has one hidden layer with 3 neurons that represent K$_p$, K$_i$ and K$_d$ that are used in traditional PID controllers. Wang et al. \cite{Wang2016ACA} proposes a double layer architecture for a non-linear system using adaptive NN control and Nonlinear Model Predictive Control (NMPC).

\subsection{Deep Neural Network Security}

Adversarial machine learning emerged when Szegedy et al.  \cite{Szegedy2013IntriguingPO} showed that the input-output mappings for \ac{DNN}s are not continuous and that it is possible to change a classification by perturbin the inputs by small amounts.
Since then, there has been a great deal of work exploring different types of adversarial attacks. Deng et al. \cite{deng2020analysis} analyze five types of adversarial attacks in autonomous driving models. 
ConAML \cite{li2020conaml} explores constrained adversarial attacks for \ac{CPS} under different settings. They propose a best effort algorithm to iteratively generate adversarial examples but don't consider the physical notion of systems for their attack model. In contrast, we introduce \attack that propagates
through a DNN's layers to produce changed outputs.
Wang et al. \cite{217595} propose ReLUVal that uses symbolic execution to show that the DNN-based systems are free from security vulnerabilities to ensure that in specific intervals, the system is free from attacks; we use MILP, which allows us to model different cost functions for different attacks; symbolic execution does not support this.

\section{Motivating Ripple Attacks: Artificial Pancreas System}
\label{attack}
\label{aps}

We present our work in the context of DNN-based \ac{APS}, closed-loop model, by Dutta et al. \cite{10.1007/978-3-319-99429-1_11}. 
The \ac{APS} model is the simplest (in terms of \ac{DNN} complexity) of our evaluation systems.
An \ac{APS} periodically administers insulin to a patient; it determines the correct dosage assuming that injections occur every $t$ minutes.

The \ac{APS} controller model has a feed-forward architecture with $2$ hidden layers of $50$ neurons each, 74 inputs and $1$ output. 
The inputs are pairs of the most recent $37$ glucose and insulin values (for a total of 74 inputs) collected at 5-minute intervals. The output is the amount of insulin to inject at the next injection time; the vast majority of the time, this value will be 0. That is, a person with an artificial pancreas doesn't require insulin every five minutes, but does require constant monitoring so that when glucose and insulin values indicate, an injection can be administered.
The controller model is trained on 30 days worth of data.
During this 30-day period, one collects all glucose/insulin samples collected over 5-minute periods as well as the amount of insulin administered at each interval. The combination of the 37 pairs of inputs and the output dosages comprise the training data set.

The system is composed of three different \ac{DNN}s.
One predicts the dose to administer in the next injection; this is the target of our adversary's attack.
The other two \ac{DNN}s are defense mechanisms designed to detect dosages below/above learned lower and upper bounds on safe injection dosage.

By design, the system will detect  an adversary who causes the system to administer a dosage outside these bounds, but it will not detect one who, for example, always administers the maximum allowed dosage. 
Thus, we model an adversary who identifies the inputs and their corresponding perturbations that will cause the \ac{APS} to administer too much insulin by repeatedly predicting a dosage just under the maximum allowed.

\subsection{Attack Model}
The attacker's goal is to manipulate sensor measurements to corrupt the output without triggering alarms.

The attacker can use network noise or physical sensor tampering to attack the systems ~\cite{10.1145/3319535.3339815}.

We assume the adversary has the following capabilities.
1) The attacker knows the precise \ac{DNN}  architecture. This information is easy to find, as the architectures are  usually specified in the  documentation. 
2) The weights and biases of the \ac{DNN}  are known as well through read-only access to the system.  
3) The attacker can modify only the inputs to the model.
4) The \ac{DNN} uses ReLU as its activation function.

\subsubsection{Strawman attacker}
A simple way to attack the system would be to change all 74 inputs to the DNN.
If all the inputs are changed, the final output prediction will undoubtedly be wrong. 
The problem with this  approach is that 
these inputs are collected every five minutes from the sensors attached to the patient, so
the attacker would have to conduct $74$ FDIAs every five minutes. 
This is impractical, so the attacker needs a better way to attack the system. 

\subsubsection{Sophisticated Attacker}
A more sophisticated approach is to perturb a small number of inputs that can
cause a change in the output. 
There are two ways to proceed:

\subsubsection{Attack 1}
The attacker can randomly choose two inputs and perturb them by large amounts. 
This too will likely cause a wrong output prediction. 
However, if the input is perturbed by a large amount, the error detection mechanisms will likely recognize it as an anomaly. 
To prevent this, the attacker can choose to perturb the two inputs by small amounts. 
However, perturbing any two random inputs by small amounts might not necessarily lead to a wrong output prediction, as shown in Section ~\ref{evaluation}. 

\subsubsection{Attack 2}
Adding one more layer of sophistication, the attacker can carefully choose the inputs to produce wrong predictions. 
This is a more targeted approach, but without knowing the precise amount by which to perturb the chosen inputs can lead to an unsuccessful attack.
If the perturbations are too high, it is likely to trigger the defense mechanism; if it is too low, it's unlikely that the output will change.

\subsubsection{Attack 3}
The ideal attack would, in addition to identifying the critical inputs, would determine the minimum perturbation that will still produce the desired change in output. 
We demonstrate that it is possible to generate this kind of attack.

\section{Challenges}
There are three challenges in designing a suitable attack mechanism. 
(1) choosing a mathematical abstraction that works best for our use case, 
(2) mapping the \ac{DNN}  model to the abstraction, and 
(3) choosing a domain-appropriate cost function for the model.

\subsection{Selecting the abstraction}
First we require a way to model the DNN that will allow identification of the critical inputs and their minimal perturbations.
There are multiple possible approaches such as \ac*{SMT} solvers, symbolic execution and \ac*{MILP}.
We seek a model that is both general and scalable.
\ac{DNN}s used in \ac{CPS} vary in size (see Table 6.1), connectivity (i.e., fully connected or not), and in terms of normalization; the air traffic management systems contain a normalization layer in the network while the \ac{APS} does not. 
The fully-connected \ac{APS} \ac{DNN} consists of $2$ hidden layers with $50$ neurons each for a total of $2800$ ($50*5 + 50*50 + 50*1$); the \ac{ACAS-Xu} consists of 5 hidden layers with 50 neurons each for $10,500$ total connections. 
Our model must be general enough to capture this range of architecture and must scale to accommodate this range of sizes. 
\ac{MILP} that satisfies both of these criteria.

\subsection{ Mapping  Mixed-Integer Linear Programming Model to Deep Neural Networks}

Transforming a DNN to a\ac{MILP} presents two challenges.  First,
as a \ac{DNN}  is composed of multiple layers, its mapping from input to outputs is a composition of the mappings at each layer.
Second, the activation functions in each layer are non-linear, and as non-linear terms are not allowed in \ac{MILP}\cite{gnonlinearity}

We address the composition of the mappings by modeling each layer as a set of interrelated linear constraints. 
We address the non-linear activation functions by decomposing them into piecewise-linear functions. We demonstrate this decomposition for the ReLU activation function, which
is used in all 3 systems we use for evaluation. 

\subsection{Choosing the cost function}
We transform the \ac{DNN} into an optimization problem that minimizes input perturbations and/or maximizes output deviations, subject to the linear constraints imposed by the \ac{DNN}'s architecture. 
The specifics of this cost function are application-dependent and depend primarily on which inputs the attacker wants to target. The cost function may naturally be limited by the perturbation bounds introduced by the attacker to avoid detection by a system's defense mechanisms, as in the case of APS.
 
The cost function can potentially introduce two problems. First,
if it is unbounded in any dimension of the feasible region, we cannot solve the \ac{MILP}.
Second, as \ac{MILP} models can be NP-hard, the the worst-case run-time can be exponential.
We demonstrate that our cost functions perform reasonably well for the three evaluation systems with time outs for some specific cases in ACAS-Xu as shown in ~\ref{evaluation}.  

\section{ReLUSyn}
\label{relusyn}
\tool is an implementation of our technique to synthesize \ac{RFDIA} attacks.
We first show how a fully connected, feed-forward \ac{DNN} can be formulated as a \ac{MILP} model. 
We use a generic representation and then show how different cost functions can be used to generate different \ac{RFDIA}.
We next show how this allows us to identify critical inputs and find minimal perturbations. 
Finally, in Section~\ref{evaluation}, we demonstrate its efficiency in three real systems. 
We also include a section on how ReLU (the non-linear activation function) is modeled as a MILP based on Setta et al. \cite{serra2018icml}  and Fischetti et al. \cite{fischetti2017deep}.


\begin{figure*}
	\centering
	\includegraphics[width=0.7\linewidth]{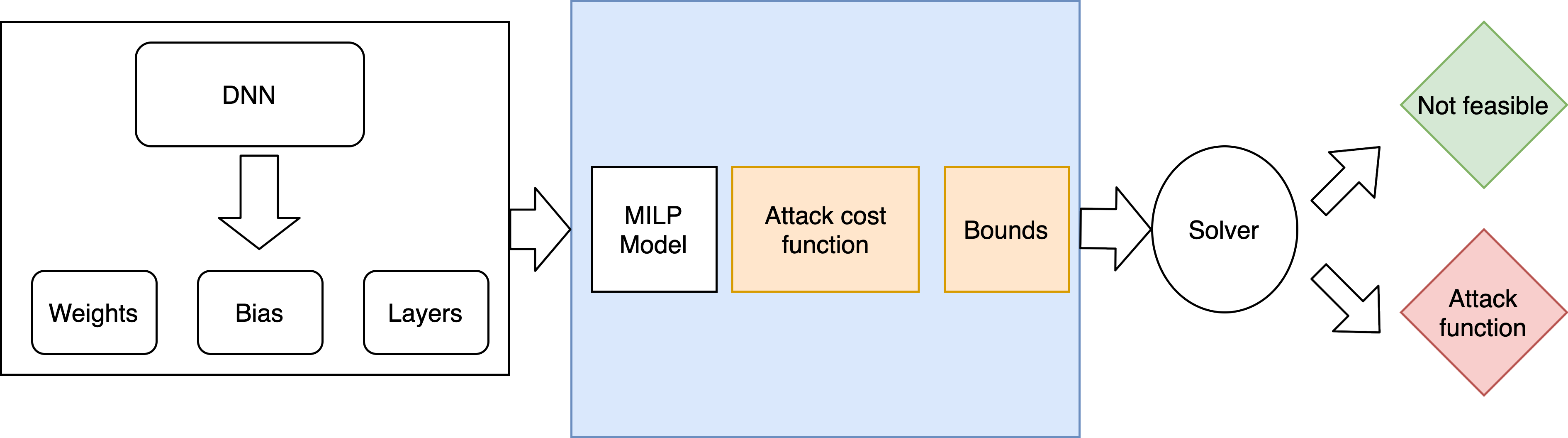}
		\caption[Methodology]{The rounded boxes depict the information provided by the users and the square boxes represent the information that is required by the solver; a MILP model is generated automatically and the cost functions and bounds are plugged in by the user.}
	\label{fig:methodology}
\end{figure*}

\begin{figure}
	\centering
	\includegraphics[width=0.7\linewidth]{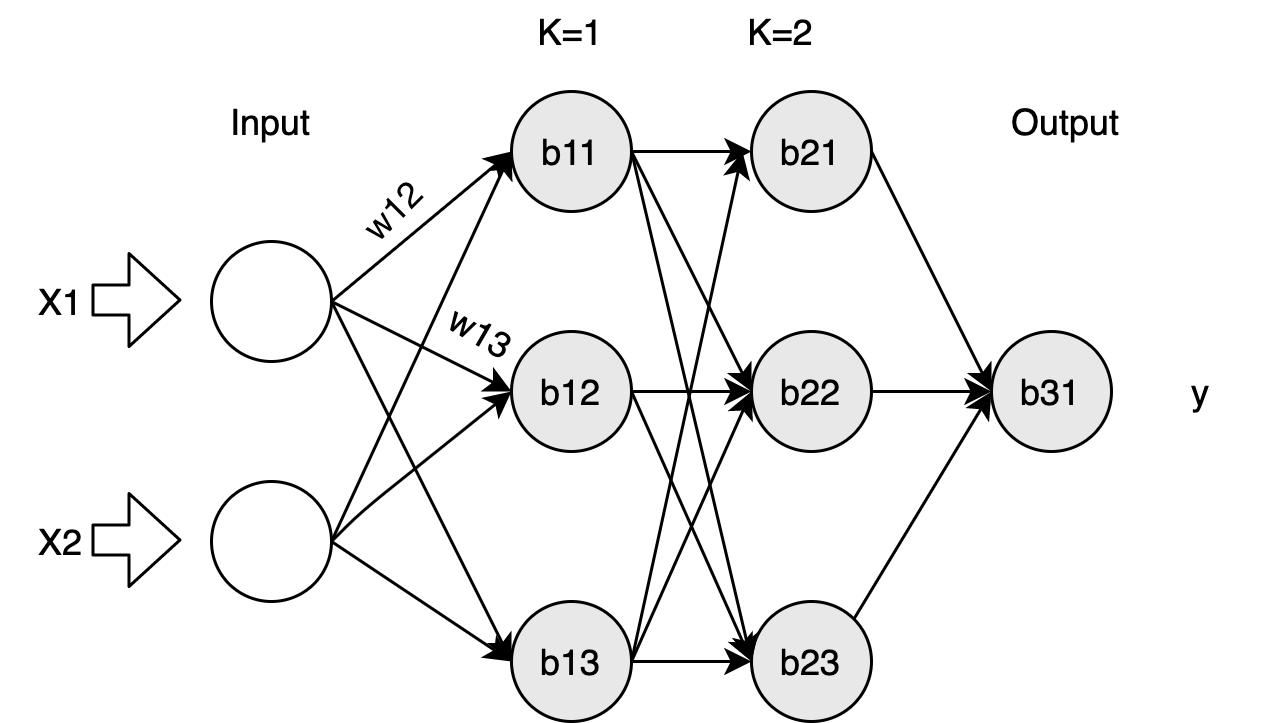}
	\caption[DNN structure]{DNN controller structure with two hidden layers K=1,2, two inputs x1 and x2, and one output y. This is an example of a fully connected network.}
	\label{fig:dnn-controller}
\end{figure}

\subsection{ReLUSyn: Overview}
Figure ~\ref{fig:methodology} depicts \tool.
\tool takes as input 1) the DNN for a CPS controller and 2) a set of bounds indicating how much the inputs can be perturbed without raising alarms. The bounds can be on inputs, outputs or both, depending on our attack model. The user can also optionally supply a custom cost function, which allows the user to request particular types of attacks (see Section \ref{section:costfunction}).
\tool then produces a \ac{MILP} such that any solution corresponds to a feasible attack.
If there is no such solution, then such no attack is possible.
	
\subsection{DNN formalism}
\begin{figure}
	\centering
	\includegraphics[width=0.7\linewidth]{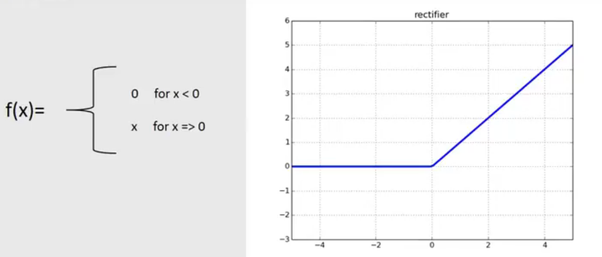}
	\caption{ReLU can be broken down into two linear units as shown in the figure for x $<$ 0 and x $\geq$ 0. This allows us to easily map the DNN equations in MILP models by breaking down the non-linearity inducing function.}
	\label{fig:relubreakdown}
\end{figure}

Consider the simple controller in Figure ~\ref{fig:dnn-controller}, which maps two inputs, $x1$ and $x2$, to a single output, $y$, via two layers.
We represent the architecture as a function F, defined as $F: X \rightarrow Y$ composed of multiple layers that map inputs X to output Y.
In our example \ac{DNN}, we call the inputs layer 0, the outputs layer 3, and have two real layers representing the hidden layers represented by $b1[1-3]$ and $b2[1-3]$.
More generally, in a
\ac{DNN} composed of $K + 1$ layers, numbered 0 to K,
layer 0 is the input and layer K is the output.
Each layer $k$ $\epsilon$ $\{1,....,K-1\}$ consists of $n_k$ nodes (neurons) (e.g., $n_1 = 3$ and $n_2=3$ in our example).
Each neuron has a bias associated with it. 
If the network is fully connected, each neuron in layer $k$ is an input to layer $k+1$. 
We denote a specific neuron by $NODE(i,k)$, which names the $ith$ node in layer $k$. 
We denote the output vector of layer $k$ as $F_k(x)$ and
refer to each output value, $NODE(i,k)$, as $F_{ik}(x)$ where $i$ $\epsilon$ $\{1,....,n_k\}$ 
For every layer $k \geq 1$ the output vector is represented by Equation ~\ref{1}, 

\begin{equation}
\label{1}
\begin{aligned}
F_k(x) &= \upsigma(W^{k-1}x^{k-1} + b^{k-1}) \\
\end{aligned}
\end{equation}

where $W$ is the weight vector, $b$ is the bias associated with each layer, $x$ is the input to the layer, and $\upsigma$ represents the activation function. 
The \ac{DNN}'s activation function determines its modeling capabilities. 
Two common ones are: rectified linear unit (ReLU) ($f(x) = max$ (${0,x}$)) as shown in Figure \ref{fig:relubreakdown} and the logistic (or sigmoid) activation function ($f(x)=1/(1+ exp(-e))$).
\tool implements ReLU, since it is one of the most commonly used activation functions (according to Krizhevsky et al. \cite{10.1145/3065386}) and is used in all of our example systems.

We restate Equation ~\ref{1} as: 

\begin{equation}
\label{2}
\begin{aligned}
F_k(x) &= ReLU(W^{k-1}x^{k-1} + b^{k-1}). \\
\end{aligned}
\end{equation}

Since a network consists of multiple layers, the general \ac{DNN} representation is a composition function where each $F_k$ represents a \ac{DNN} layer. 
\begin{equation}
\label{3}
	\begin{aligned}
	F(x) &= F_K \circ F_{K-1} \circ F_{K-2} ....... \circ F_1(x),    \\
	or \\
	F(x) &= F_K ( F_{K-1}( F_{K-2} .......  (F_1(x)))),    \\
	\end{aligned}
\end{equation}

\label{section:attacks}

\begin{figure}
	\centering
	\includegraphics[width=0.7\linewidth]{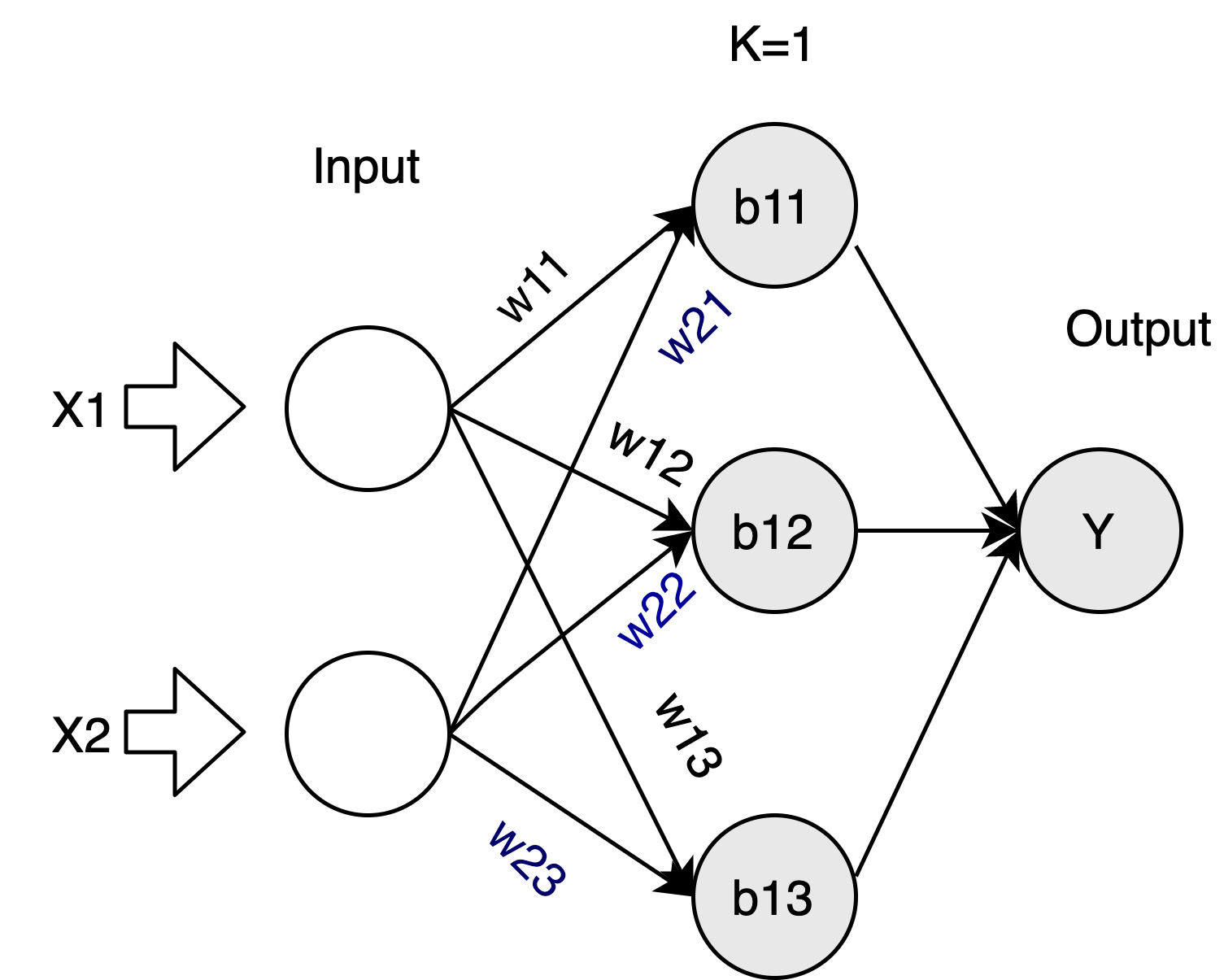}
	\caption[APS]{APS that takes in two inputs which we consider as the sensor values from the human. It predicts the amount of insulin to be injected at some time based on the sensor inputs.}
	\label{fig:toyaps}
\end{figure}

\subsection{Building MILP model \& attack-specific cost functions}
\label{section:costfunction}
In theory, we need to decompose our activation function, $f(x) = \max(0, x)$, into two piecewise linear functions to satisfy the \ac{MILP} requirements.
However, we use the Gurobi solver~\cite{gurobi}, which offers direct support for $\max$, thus Algorithm \ref{algo:c}, which describes how to transform a \ac{DNN} into a \ac{MILP}, uses ReLU directly.

Now that we have formulated the DNN as a MILP model, we next specify a cost or objective function, which allows the user to specify the costs of perturbing specific inputs; the perturbation bounds are determined by the lower and upper bounds specified for these variables in the model, providing the attacker greater flexibility if, for example, some inputs are easier to perturb than others.

As discussed in Section \ref{attack}, the attacker's goal is to change the inputs to produce a desired output change without triggering an alarm.
More formally, the attacker's goal is to change the output $y$ to $y'$, where $y' = y + a$ for some constant $a$. 



In our use cases, we use cost functions that express both random and targeted \ac{RFDIA}, where the goal is to minimize the input deviations, with or without limiting the set of variables that can be perturbed.
We model \ac{RFDIA} by introducing a new variable, $\Delta x$, the amount by which an input can be perturbed, which we then minimize.
The objective can also include coefficients per input, to reflect, for example that some inputs are most difficult to perturb or that the attacker finds some inputs more attractive perturbation targets.
The procedure for maximizing output deviations is similar, except we do not require a corresponding deviation variable. The output variable is simply added to the MILP model and set equal to the result of the final DNN layer; this variable can then be included in the objective function (hierarchical or otherwise) to maximize or minimize the DNN output.

Hence, in our final formulation, Equation \ref{1} is rewritten as Equation \ref{11}, 

\begin{align}
\label{11}
y &=  ReLU(W(x + \Delta x ) + b).
\end{align}

Algorithm \ref{algo:c} shows how to incorporate the \ac{RFDIA} cost functions and build the MILP model. 
 
\begin{algorithm}
	Inputs:input ($x$), weight matrices, bias vectors, num\_layers, num\_neurons
	Outputs:input\_delta ($\Delta x$), layer\_output, ReLU\_output
	
	\textbf{Model}, \linebreak
	(I) \textcolor{red}{For the first layer $k = 0$ (with weight matrix $W_0$ and bias vector $b_0$),} $\color{red} output_0 = W_0 * (x + \Delta x) + b_k$
	\linebreak
	(II) For every subsequent layer $k \geq 1$ (with weight matrix $W_k$ and bias vector $b_k$), $ output_k = W_k * input_k + b_k$
	\linebreak
	(III) Apply activation function to every layer's output:
	\linebreak
	Constraints: $ReLU\_output_k = max(0, output_k)$ (component-wise) \
	\linebreak
	(IV) \textcolor{red} {Apply constraints to the DNN output to force the desired deviation}
	\linebreak
	The final layer's rectified output is the DNN output.
	
	\textbf{Cost/Attack Function} \linebreak
	$ \color{red} \min |\Delta x|$
	\caption{Modeling neural network in MILP with perturbation variables and a cost function}
	\label{algo:c}
\end{algorithm}

An attacker can choose which inputs to attack by assigning a coefficient of $0$ to those inputs perturbation variables where the inputs that can be freely perturbed and larger coefficients for those variables where the inputs should not be modified.
This is crucial in \ac{CPS} where inputs may come from different sensors and the attacker may be able to tamper with some, but not all, of them.

Considering different scenarios, the attacker can minimize the values depending on which inputs they are interested in targeting with an \ac{RFDIA}. 
The \ac{RFDIA} cost function in Algorithm ~\ref{algo:c} minimizes the absolute values of the perturbations to both inputs.

\subsection{Synthesizing attacks}
Once the attacker has a model and the cost function, which will assign low costs to perturbing critical inputs, designed, the next step is to determine the desired perturbations.
In the running example of the \ac{APS}, the inputs consist of the 74 readings from the two sensors collected every five minutes.
The output is the amount the insulin to be injected, subject to the constraint that we minimize the change in inputs.
The solver produces a set of inputs which, upon precise perturbation, will cause an undetectable change in the dose of insulin to be administered.
Given the specific inputs and their perturbations,
the attacker can successfully mount the attack.

\section{Evaluation}
\label{evaluation}

Table \ref{tab:table1} presents the architectures of the three \ac{DNN}s used in the \ac{APS}, \ac{HCAS}, and \ac{ACAS-Xu}, in increasing order of complexity. 
These applications are all mission critical and can produce catastrophic outcomes in the presence of a successful attack.
They also come from two different domains: medical devices and autonomous vehicles, providing evidence of the generality of \tool.

\begin{table}[ht!]
	\begin{center}
		\caption{System descriptions}
		\label{tab:table1}
		\begin{tabular}{l|r|r|r}
			\textbf{} & \textbf{APS} & \textbf{HCAS} & \textbf{ACAS-Xu} \\
			\hline
			\textbf{Size of the networks} &  &  &  \\
			Number of inputs &  74&   5&  5\\
			Number of hidden layers &  2&  5&  6\\
			Neurons/Layer &  8&  25 & 50 \\
			Number of outputs & 1&  3& 5\\
			\hline
			\hline

		\end{tabular}
	\end{center}
\end{table}

\tool accepts a trained \ac{DNN} in TensorFlow \ac{DNN} format \cite{TensorFlow}.
It then transforms the \ac{DNN} to a \ac{MILP} using code we developed in Python.
Finally, we use Gurobi to solve the \ac{MILP} to produce the set of critical inputs and their minimal perturbations.
If Gurobi cannot find a feasible solution to the \ac{MILP}, that indictes that no attack for the given objective function is possible.

We ran our experiments on a server with a 2.40 GHz (10 MB L3 cache) Intel Xeon E5-2407 v2 processor
with 8 cores across 2 NUMA nodes and hyperthreading disabled. The server uses Ubuntu 16.04.6 LTS with 32 GB RAM that
is uniformly distributed (16 GB each) across both NUMA nodes.

\subsection{Case study: APS}

The \ac{APS} takes 74 inputs, $37$ pairs of insulin and glucose values gathered every $5$ minutes, and produces as output a predicted insulin dosage.
In under a minute, \tool successfully synthesizes attacks that identify which insulin and/or the blood glucose values to perturb and the amounts by which to perturb them to elicit the maximum output change, subject to the constraints for a given test.

\begin{table}[]

\caption{Targeted Attacks - 1 input perturbed at a time}
\label{1inputAPS}
\resizebox{\columnwidth}{!}{
\begin{tabular}{lllll}
\hline
\multicolumn{1}{|l|}{Output Range} & \multicolumn{1}{l|}{Attacks (Successful/Total)} & \multicolumn{1}{l|}{Peak attack time (s)} \\ \hline
\multicolumn{1}{|l|}{210-215} & \multicolumn{1}{l|}{0/74} & \multicolumn{1}{l|}{N/A} \\ \hline
\multicolumn{1}{|l|}{215-220} & \multicolumn{1}{l|}{0/74} & \multicolumn{1}{l|}{N/A} \\ \hline
\multicolumn{1}{|l|}{220-221} & \multicolumn{1}{l|}{0/74} & \multicolumn{1}{l|}{N/A} \\ \hline
\end{tabular}
}
\end{table}

\begin{table}[]

\caption{Targeted Attacks - 2 inputs perturbed at a time}
\label{2inputAPS}
\resizebox{\columnwidth}{!}{
\begin{tabular}{lllll}
\hline
\multicolumn{1}{|l|}{Output Range} & \multicolumn{1}{l|}{Attacks (Successful/Total)} & \multicolumn{1}{l|}{Peak attack time (s)} \\ \hline
\multicolumn{1}{|l|}{210-215} & \multicolumn{1}{l|}{13/2701} & \multicolumn{1}{l|}{0.006} \\ \hline
\multicolumn{1}{|l|}{215-220} & \multicolumn{1}{l|}{0/2701} & \multicolumn{1}{l|}{N/A} \\ \hline
\multicolumn{1}{|l|}{220-221} & \multicolumn{1}{l|}{0/2701} & \multicolumn{1}{l|}{N/A} \\ \hline
\end{tabular}
}
\end{table}

\begin{table}[]

\caption{Targeted Attacks - 74 inputs perturbed at a time}
\label{74inputAPS}
\resizebox{\columnwidth}{!}{
\begin{tabular}{lllll}
\hline
\multicolumn{1}{|l|}{Output Range} & \multicolumn{1}{l|}{Attacks (Successful/Total)} & \multicolumn{1}{l|}{Peak attack time (s)} \\ \hline
\multicolumn{1}{|l|}{204-210} & \multicolumn{1}{l|}{1/1} & \multicolumn{1}{l|}{0.06776785851} \\ \hline
\multicolumn{1}{|l|}{210-215} & \multicolumn{1}{l|}{1/1} & \multicolumn{1}{l|}{0.080} \\ \hline
\multicolumn{1}{|l|}{215-220} & \multicolumn{1}{l|}{1/1} & \multicolumn{1}{l|}{0.085} \\ \hline
\multicolumn{1}{|l|}{220-221} & \multicolumn{1}{l|}{1/1} & \multicolumn{1}{l|}{0.097} \\ \hline
\end{tabular}
}
\end{table}

\begin{table}
\caption{Random attacks on 201 different input value sets, given a specified input perturbation bounds and minimum output perturbation threshold. 
This input perturbation bound is $[\max(-x_{i}, -20), 20]$ for the former and $[\max(-x_{i}, -5), 5]$ for the latter two attacks respectively and the minimum output perturbation threshold is $+10$ for the former two attacks and $+20$ for the latter attack respectively.}
\label{APSrandom}
\resizebox{\columnwidth}{!}{
\begin{tabular}{lllll}
\hline
\multicolumn{1}{|l|}{Inputs Perturbed}& \multicolumn{1}{|l|}{Attacks (Successful/Total)} &  \multicolumn{1}{l|}{Peak attack time (s)} \\ \hline
\multicolumn{1}{|l|}{1} & \multicolumn{1}{|l|}{293/14874} & \multicolumn{1}{l|}{0.006} \\ \hline
\multicolumn{1}{|l|}{2} & \multicolumn{1}{|l|}{2369/542901} & \multicolumn{1}{l|}{0.010} \\ \hline
\multicolumn{1}{|l|}{74} & \multicolumn{1}{|l|}{201/201} & \multicolumn{1}{l|}{0.015} \\ \hline
\end{tabular}
}
\end{table}

\subsection{Experimental methodology}

Our experimental methodology is as follows:
given a set of accurate inputs, we construct \emph{perturbation scenarios}.
A perturbation scenario consists of the number of inputs to perturb and bounds on the maximum absolute amount by which to perturb them.
For each perturbation scenario, we evaluate our attack efficacy for different outcome constraints.
We produce two kinds of attacks, targeted and random.
Our targeted attacks all use a single input/output set, where the correct dose is $204$.
In these attacks, our goal is to maximize the output, subject to constraints on the range of allowed outputs.
Tables ~\ref{1inputAPS}, ~\ref{2inputAPS}, ~\ref{74inputAPS} present results for these attacks.
The random attack scenario maximizes the DNN output subject to the constraint that the new output is at least some threshold higher than the correct output; input perturbation bounds apply in this case as well.
Thus, if $a$ is the actual dosage, $m$ is the minimum change required to consider an attack successful, and $UB$ is the upper bound of the normal range of doses, the random attack scenario corresponds to a targeted attack scenario with output range of $[a + m, UB]$.

Tables \ref{1inputAPS}, \ref{2inputAPS}, and \ref{74inputAPS} provide detailed results for the targeted attack scenarios with one, two, and 74 perturbed inputs, respectively.
For the single input case (Table \ref{1inputAPS}), we iterate over the 74 different inputs, allowing each input to be perturbed by an amount in the range $[\max(-x_{i}, -5), 5]$ (to ensure that our perturbations cannot drive the input below 0).
Although we formulate the perturbation bounds generally, we add domain specific constraints to the formulation. In particular, blood glucose readings fall in the range $0 - 0.3$ mg/dL. Inducing a change of $+5$ mg/dL would produce invalid inputs, so we constrain inputs for these variables to fall within the $0-0.3$ range.
The insulin values are usually in the range $100-300$, so the perturbation bounds apply to those inputs.

The multiple input scenarios work the same way, but produce a different number of total possible attacks, i.e., $(74*73)/2$ for the 2-input case, and just 1 for the 74-input case. The Gurobi model then tries to find the smallest perturbation to the chosen input that produces the maximal output within the specified range. The Gurobi model has a multi-hierarchical objective function that focuses first on maximizing the output change and then minimizing the absolute values of the input perturbations without degrading the first objective. This allows the attacker to use \tool as a black box system. 

Table ~\ref{APSrandom} shows the results for random attacks, in which we consider many input/output pairs, different input range perturbations, and different minimum required output changes. We select 201 random input/output pairs and create attack scenarios for 1-input, 2-input, and 74-input perturbations.
For the single input scenario, we use an input perturbation range of  $[\max(-x_{i}, -20), 20]$ and consider an attack successful only if it increases the correct dosage by at least 10 (a typical dosage ranges between 100 and 300).
For the two-input scenario, we use a smaller input perturbation range of $[\max(-x_{i}, -5), 5]$ with the same output threshold, and for the 74-input scenario, we use the same input perturbation range, but require an output change of at least 20.

\subsection{Result Analysis}
It is difficult to attack the system with the heavy constraints that we impose on the targeted attacks.
In fact, there are no single-input attacks for the given output ranges.
Even in the 2-input scenarios, we can only find attacks in the range closest to the actual dosage (this range corresponds to an overdose between 6 and 11 mg/dL).
We also explored scenarios with 4, 5, and 6 perturbations and find that the more perturbations allowed, the easier it is to find an attack.
Unsurprisingly, when we can change all the inputs (the 74-input scenario), we can always produce an attack.
More importantly, the time to find (or not find) an attack is sufficiently short (60 ms) that it is possible to identify the right combination of only two inputs to perturb in under 17 seconds. This means that it is entirely feasible to launch a constant barrage of such attacks, leading to a potentially lethal overdose by way of many incremental overdoses.

The results for the random attacks are even more worrisome. We can elicit output changes of 10 or 20 mg/dL relatively easily in all the scenarios.
This techniques is so performant, that an attacker can easily launch a successful \ac{RFDIA}. In many cases, perturbing only a single input by a small amount (e.g., $[-5,5]$) is enough to cause harm.
As we will see in the following sections, the aviation systems are equally vulnerable.

\section{Case Study: ACAS-Xu and HCAS}

\begin{table}[]
		\caption{HCAS Results: Each column presents results for three different attack scenarios, perturbing 1, 2, or 3 (of the 5) inputs. Combinations reports the  total number of attacks for each scenario; Successful attacks indicates how many of those attacks were successful. The peak time is the maximum time for any attack, with an X indicating that the experiment did not complete in 7 hours.}
		\label{hcas}
\begin{tabular}{|l|l|l|l|}
\hline
\begin{tabular}[c]{@{}l@{}}Combinations\\ (1A, 1B, 1C)\end{tabular}       & \begin{tabular}[c]{@{}l@{}}Successful \\ attacks\end{tabular}      & Peak time (s)               & Memory used            \\ \hline
\multicolumn{4}{|l|}{Random Attack; minimize 1 output; max perturbation: 5}                                                                                \\ \hline
12, 12, 4                                                                 & 5, 11, 4                & 5.7, 94.3, 80.9         & 70.2, 108, 103.5       \\ \hline
\multicolumn{4}{|l|}{\begin{tabular}[c]{@{}l@{}}Fully-targeted Attack, no maximum perturbation\end{tabular}} \\ \hline
3, 3, 1                                                                   & 0, 0, 0                 & 7.9,174.4,1130.6      & 3.1, 114.9, 152.7      \\ \hline
\multicolumn{4}{|l|}{Partially-targeted Attack}                                                               \\ \hline
3, 3, 1                                                                   & 2, 3, 1                 & 3.0, 4.2, 2.7           & 69.0, 71.5, 67.4       \\ \hline
3, 3, 1                                                                   & 2, 3, 1                 & 5.2, 38.9, 5.1          & 69.4, 83.1, 76         \\ \hline
3, 3, 1                                                                   & 0, 0, 1                 & 4.9, 229.6, 1.8         & 69.7, 110.4, 58.9      \\ \hline
3, 3, 1                                                                   & 0, 0, 1                 & X, X, 357.4             & X, X, 111.1            \\ \hline
\end{tabular}
\end{table}

\begin{table}[]
	\caption{ACAS-Xu Results: Each column presents results for three different attack scenarios, perturbing 1, 2, or 3 (of the 5) inputs. Combinations reports the  total number of attacks for each scenario; Successful attacks indicates how many of those attacks were successful. The peak time is the maximum time for any attack, with an X indicating that the experiment did not complete in 7 hours.}
	\label{acasxu}
\begin{tabular}{|l|l|l|l|}
\hline
\begin{tabular}[c]{@{}l@{}}Combinations\\ (1A, 1B, 1C,)\end{tabular} & \begin{tabular}[c]{@{}l@{}}Successful \\ attacks\end{tabular} & Peak time (s)         & \begin{tabular}[c]{@{}l@{}}Memory used\\ /MB\end{tabular} \\ \hline
\multicolumn{4}{|l|}{Random Attack; 1 output minimized; Delta range: 5}                                                                                                                                                  \\ \hline
20, 40, 40                                                           & 2, 9, 12                                                      & 465.8, 4141.1, 6644.9 & 160.7, 209.1, 269.1                                       \\ \hline
\multicolumn{4}{|l|}{\begin{tabular}[c]{@{}l@{}}Targeted Attack; complete ordering on outputs; \\ Deltas are restricted at 5\end{tabular}}                                                                               \\ \hline
5, 10, 10                                                            & 0, 0, 0                                                       & 739.4, 3678.1, 7296.6 & 204.8, 202.6, 296                                         \\ \hline
\end{tabular}
\end{table}

The ACAS Xu and HCAS are both examples of aircraft collision avoidance systems, but they have rather different network architectures, as shown in Table ~\ref{tab:table1}.

\begin{figure}
	\centering
	\includegraphics[width=0.7\linewidth]{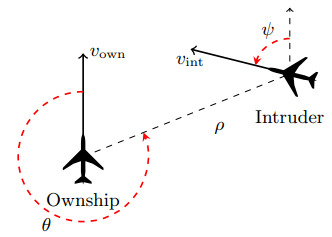}
	\caption[ACAS Xu]{ACAS Xu geometry}
	\label{fig:acasxugeometry}
\end{figure}

Figure ~\ref{fig:acasxugeometry} shows the geometry of \ac{ACAS-Xu}. 
The \emph{ownship} is the unmanned vehicle with the \ac{CA} system and the attacker's goal is to crash the \emph{intruder} into the ownship. 
The \ac{ACAS-Xu} models a large lookup table mapping sensor readings to navigation decisions.
This mapping is represented by forty-five different \ac{DNN} models. 
Each network has the architecture described Table \ref{tab:table1} with $5$ inputs, $6$ hidden layers and $5$ outputs.  

There are five sensor readings (inputs to the network): 
\begin{enumerate}
	\item Distance from ownship to intruder ($\upvarrho$)
	\item Angle to the intruder relative to ownship heading direction ($\psi$)
	\item Header angle to intruder relative to ownship heading direction  ($\theta$)
	\item Speed of ownship ($v_{own}$)
	\item Speed of intruder($v_{int}$)
\end{enumerate}

The $5$ outputs are:
\begin{enumerate}
	\item clear-of-conflict (COC): keep going in the same direction
	\item weak right: Turn ($1.5\degree /s$) to right
	\item strong right:  Turn  ($3 \degree /s$) to right
	\item weak left: Turn ($1.5\degree /s$) to left
	\item strong left: Turn ($3 \degree /s$) to left
\end{enumerate}

The \ac{ACAS-Xu}'s goal is to avoid crashes between the ownship and an intruder.
If the two are too close, the \ac{ACAS-Xu} output should be an action that moves them apart. 
The output is expressed as a score for each of the five possible actions, and the best action is the one with
the minimal score. 

The \ac{HCAS} functions similarly to \ac{ACAS-Xu}.

We conduct two types of \ac{RFDIA}: random, partially-targeted, and fully-targeted.
In a random attack, the attacker requires only a partial order -- that a particular action is given the lowest score.
In a partially targeted attack, the attack designates, in order, the two actions with minimal scores. For example, if the correct action is \emph{strong right}, the attack might want an ordering with \emph{strong left} followed by \emph{weak left}.
In a full-targeted attack,
the attacker requires a total ordering on the outputs.

\subsection{Experimental Methodology}
We consider a set of accurate inputs and construct perturbation scenarios. A perturbation scenario consists of 
the set of inputs to be perturbed and the output is an ordering of the possible actions. A targeted attack requires that a specific action have the smallest value; a random attack considers as successful any ordering whose minimum-valued action is not the correct action.

\subsubsection{Random Ordering Attack}
Unlike the \ac{APS}, which produces a single value as output, both the \ac{ACAS-Xu} and \ac{HCAS} produce scores whose relative values imply an action.
In a random attack, if the normal output was $strong$ $right$ the attacker's goal might be to
to change the output to $COC$, so that the ownship would take no action to avoid the intruder.
We express this desired outcome by adding a constraint to the MILP that ensures that the score of $COC$ is the least among all other outputs. 
In a random attack, this is sufficient, as the attacker's goal is only to manipulate the lowest scoring action.

For each perturbation scenario, we evaluate attack efficacy for different orderings, which means that if the correct decision based on the scores is $1$,  we perturb the inputs such that the output changes to $2$ (i.e., so that the score for action $2$ is smaller than that for all other actions).

The first set of results in Tables \ref{hcas} and \ref{acasxu} is an example of a random attack by perturbing $1$, $2$ or $3$ of the inputs.
A random attack is attractive for an attacker, because
there are only four constraints (that the desired output has a value lower than each of the others), and the attack is sufficiently fast that the attacker can easily launch repeated attacks while the intrude and ownship are in the same vicinity.

\subsubsection{Targeted Attacks}
 
The targeted attacks more heavily constrain the outputs.
Both Tables \ref{hcas} and \ref{acasxu} present results for the fully-targeted attack, where the attacker selects a complete ordering on the outputs.
It is possible that the attack can better hide the attack by constructing a total ordering without anomalies, such as the two minimal scores providing contradictory information such as \emph{strong left} and then \emph{weak right}.
Table \ref{hcas} additionally present three partially-targeted attacks, where the attacker specifies only the actions receiving the two lowest scores.

\subsection{Result Analysis}
Although these \ac{DNN}s are much more complicated that than in the \ac{APS}, it is still relatively easy to launch an \ac{RFDIA}. In fact,we found successful single-input perturbation attacks where perturbing a single input was enough to change the output decision.
However, when the attacker required a complete ordering, we were unable to find a single successful attack.
This is not terribly surprising, as the problem is over-constrained.
Although some of our experiments timed out, and others took many minutes, there were enough successful attacks that completed in only a few seconds to indicate that \ac{RFDIA} are legitimate attacks requiring mitigation.

\section{Result Summary}
Table ~\ref{summary} presents a summary of all three systems.

\tool is effective at finding attacks on different systems, and in many cases, does so quickly enough to cause concern for the security of these systems.
We conducted ~555,000 attacks for the \ac{APS}, of which ~5500 were successful in under one second per attack.
We conducted ~30 \ac{HCAS} random attacks out of which most (~20) were successful, and the maximum amount of time taken by the attacks was 41 seconds.
Targeted attacks are significantly more difficult to launch, of the ~7 targeted attacks we tried for \ac{HCAS} only 1  was successful.
The results for \ac{ACAS-Xu} were similar, albeit slower.
Of ~140 random attacks, ~32 were successful, and the maximum amount of time taken by the attacks was 7228 seconds;
of the ~16 targeted attacks only 2  were successful.

Thus,  synthesizing \ac{RFDIA} attacks is both feasible and practical.
In many cases, perturbing only a single input can significantly affect the output.

 \begin{table}[ht!]
 	\centering
 	\caption{Results Summary}
 	\label{summary}
 	\begin{tabular}{c|c|c|c}
 	
 		\textbf{Inputs-pert} & \textbf{Total} &  \textbf{Successful} & \textbf{Mean-Time/ } \\
 		urbed/Attack &  Attacks &  Attacks &  Attack(s) \\
 		\hline
 		\multicolumn{4}{c}{APS}\\
 		\hline
 		1 &  15170 & 342 &  <1\\
 		2 &  553705 & 4944 &  <1\\
 		74 &  205 & 205 & <1\\
 		\hline
 		\multicolumn{4}{c}{HCAS - Random}\\
 		\hline
 		1 &  12& 5 &  3.87\\
 		2 &  12& 11 &  32.89\\
 		3 &  4&  4&  41.14\\
 		\hline
 		\multicolumn{4}{c}{HCAS - Targeted (Partial Ordering)}\\
 		\hline
 		1 &  3& 1&  2.14\\
 		2 &  3& 0 &  2.91\\
 		3 &  1&  0&  2.74\\
 		\hline
 		\multicolumn{4}{c}{ACAS Xu - Random}\\
 		\hline
 		1 &  20& 2&  100.11\\
 		2 &  40& 9& 548.72\\
 		3 &  40& 12&  1452.176\\
 		4 &  20&  8&  5029.58\\
 		5 &  20&  2&  7228.966\\
 		\hline
 		\multicolumn{4}{c}{ACAS Xu- Targeted (Partial Ordering)}\\ 
 		\hline
 		1 &  5& 0&  297.2\\
 		2 &  10& 1 &  23595.58\\
 		5 &  1&  1&  8311.21\\
 		\hline
 		\hline
 	\end{tabular}
 \end{table}

\section{Designing Robust DNNs}
Our experiments showed that it was significantly easier to launch \ac{RFDIA} on the \ac{APS} than on either \ac{HCAS} or \ac{ACAS-Xu}.
Although the \ac{APS} and \ac{ACAS-Xu} are
both fully connected, feed-forward architectures, they differ in two fundamental ways: 1) the \ac{APS} has only two hidden layers, while \ac{ACAS-Xu} has 6, and 2) \ac{APS} has no normalization layer.
We hypothesize that these differences account for the difference in difficulty launching \ac{RFDIA}.

The six layers of \ac{ACAS-Xu} means that the ripples that propagate from the inputs become diluted as they progress through the layers.
This both increases runtime for the attacks and makes it more difficult to find the right combination and magnitude of the perturbations.

The normalization layer effectively
masks the small perturbations of the sensor inputs from the DNN layers, making it more difficult to successfully launch an attack, because we impose bounds on the perturbations to preserve MILP scalability (as well as to avoid detection).

.

\section{Conclusion}
\label{conclusion}
We have introduced a new kind of attack, the \ac{RFDIA}, which is both feasible and practical to launch against \ac{DNN}-based CPS controllers.
We demonstrate \tool, which finds attacks on difference classes of \ac{DNN}s in second.

There are two limitations  \tool that suggest avenues for future work.
First, \tool assumes that the attacker has read access to the weights and the bias of a \ac{DNN}.
Even in an ideal scenario, it is difficult to obtain read access to the weights and the bias of safety-critical systems such as \ac{CA} systems since these are often hidden.
Therefore, the attacker would need tools or techniques that allows them to attack the system without having knowledge about the internals of the network.

Second, for complex \ac{DNN}s our tool modeled, there were many cases when the solver timed out, and did not produce any attacks. 
Timeouts occur because the search space for the MILP model and given input was too large.
It is possible that identifying important inputs, choosing appropriate perturbation bounds, and selecting a suitable starting point would limit the search space and significantly scale our approach.

	\label{section:limitations}

We also have identified three extensions that would greatly improve \tool.
First, it can be extended for different application domains to find attacks.
For example, \ac{AV} such as self-driving cars have multiple stacks of \ac{DNN}s to compute outputs, and \tool can be utilized to find attacks in these chained \ac{DNN}s. 
This requires extending the MILP model-building phase to allow multiple \ac{DNN}s to be analyzed together and be tested for \ac{RFDIA} or any other attack.
Second, one can model cost functions for an attack model different from ours and use our insights to generate attacks. This should be relatively straight forward due to the \ac{DNN} modeling and the fact that \tool can easily incorporate different attack specific cost functions/MILP model modifications. 
Currently, there are a lot of unexplored vulnerabilities in \ac{DNN}-based \ac*{CPS}; thus one can use \tool to model new vulnerabilities. 
Finally, we can use \tool to debug existing DNNs.
In our work, we proposed \tool to generate attacks for trained \ac{DNN}s that also tell us if in a particular setting, no attacks exist. 
However, instead of trying to find attacks in the \ac{DNN}, we can also use \tool for debugging; it means that we can conduct an input-output pair generation to study if there are unexpected output values for input vectors not accounted for in the training phase of the DNN.

\bibliographystyle{ACM-Reference-Format}

\bibliography{sample-base}
\newpage

\appendix

\section{MILP Model}
The non-linear ReLU activation function cannot be modeled directly by MILP solvers, so we present an equivalent formulation, which we use to identify critical inputs and find the minimum perturbations, as discussed in \cite{fischetti2017deep}.

 To create a \ac{MILP} model, we study the basic scalar equation that describes the \ac{DNN} architecture where $w$ is the weight, $y$ is the input and $b$ is the bias. 

\begin{equation}
\label{4}
\begin{aligned}
x &= ReLU(w^Ty + b) \\
\end{aligned}
\end{equation}

We cannot apply the $f(x) = \max(0, x)$ operator directly because it is non-linear. 
As explained in Figure \ref{fig:relubreakdown}, the ReLU function is piecewise linear and can be broken into two parts, namely the positive and negative halves of the domain.

In theory $f(x)$ = $max(0,x)$ can be represented as Equation ~\ref{12}, however, \ac{MILP} (as the name implies) does not support non-linear constraints and piecewise linear functions must be modeled with creativity.
Hence to model ReLU in \ac{MILP}, we introduce a dummy variable $s$.

\begin{equation}
\label{12}
\begin{aligned}
& if (x < 0) \\
& & then(0); \\
& else (x);	 \\
\end{aligned}
\end{equation}

We can rewrite Equation ~\ref{4} given above as follows:

\begin{equation}
\label{5}
\begin{aligned}
w^Ty + b = x - s, x \geq 0, s \geq 0 \\
\end{aligned}
\end{equation}

The addition of the variable $s$ in Equation ~\ref{5} maintains linearity while modeling ReLU functionality, as $x$ will always be non-negative.

Equation ~\ref{5} by itself is problematic as there are infinitely many $x, s \ge 0$ that would satisfy this equation if there is a solution.
For example, if $x', s'$ is a possible solution, then so is $x' + 5, s' + 5$ and so on. 
This is why we need to introduce another variable $ac \in \{0, 1\}$ which adds a separate constraint to the model depending on its value.

\begin{equation}
\label{6}
\begin{aligned}
ac =  1 \rightarrow x \leq 0  \\
ac =  0 \rightarrow s \leq 0  \\
ac \in \{0,1\} \\
\end{aligned}
\end{equation}

If $ac = 1$, this will add the constraint $x \leq 0$ to the model. This coupled with $x \geq 0$ will mean $x$ will be forced to be $0$ and $Wy + b = - s \leq 0$ (but $x$ which represents the output to the next layer would be $0$).

If $ac = 0$, this will add the constraint $s \leq 0$ to the model. This coupled with $s \geq 0$ will mean $s$ will be forced to be $0$ and $Wy + b = x \geq 0$ (and $x$ which represents the output to the next layer would be $\geq 0$).

Adding regularization to $ac$ will mean there is a unique solution for $x$ and $s$. This regularization will try to clamp $ac$ to 0, which will force $s$ to be 0 and consequently force the result of the LHS to be greater than or equal to 0. The main takeaway is that it is possible to model ReLU(x) (or max(0, x)) internally in a MILP solver.

Modern MILP solvers can handle these constraints directly.

Extending this scalar example to the DNN case leads to a \ac{MILP} model of the form shown below.

We have divided our \ac{MILP} model into three main parts.

Equation ~\ref{7} represents the cost function that is to be minimized or maximized.
This cost function is different for different applications and depend on the goal; for an attack synthesis application, minimization of input deviations is the goal. Different inputs can be weighed differently and the cost function can be adjusted by penalizing the perturbations to certain inputs less (through smaller coefficients).
The equation \ref{7} is a generic representation of cost function modeling. 
We explain our cost function in more detail in \ref{section:costfunction} for \ac{RFDIA} attack synthesis. 

\begin{equation}
\label{7}
\begin{aligned}
& \underset{}{\text{min/max}}
& & f(\Delta x) \\
\end{aligned}
\end{equation}

Equations ~\ref{8} represents the ReLU modeling for each layer in the network. 
We showed in Equation \ref{6} how we can represent ReLU units as a set of linear constraints in \ac{MILP} solvers. 
To model the constraints, we use the formalism from Equation \ref{5} where we introduce the variable $s$ to accommodate the non-linearity using linear representations as explained in (\ref{6}). 
We expand this and apply it to all the layers in the modeling.

The model constraints appear below (mainly modeling ReLU and setting inputs to DNN layers appropriately):
\begin{equation}
\label{8}
\begin{aligned}
& \underset{}{\text{min/max}}
& & f(\Delta x) \\
& \text{subject to} & &  \sum_{j=0}^{n_k} w_{ij}^{k-1}x_{j}^{k-1} + b_i^{k-1} = x_i^k - s_i^k  \\
& & & x_i^k \geq 0, \\
& & & s_i^k \geq 0 \\
& & & ac_i^k  \epsilon  \{0,1\} \\
& & & ac_i^k  =  1 \rightarrow  x_i^k \leq 0  \\
& & & ac_i^k =  0 \rightarrow s_i^k  \leq 0   \\
\end{aligned}
\end{equation}

Equations ~\ref{9} adds lower bounds ($lb$) and upper bounds ($ub$) in the model for input variables to the first DNN layer. We are essentially limiting how much we can perturb the inputs by through these bounds. This will limit our search space and make the MILP model more tractable to solve. The output might also be bounded depending on what the attacker wants to achieve. All other variables in our model are unbounded (other than being non-negative due to the ReLU activation).
 
\begin{equation}
\label{9}
\begin{aligned}
& & & lb \leq x_i^0 \leq ub, \\
\end{aligned}
\end{equation}

Our MILP modeling applies the $f(x) = \max(0, x)$ operation directly to the output of layers because the Gurobi solver linearizes this internally when solving the model. It offers direct support for the $\max$ function. This formulation may or may not be how Gurobi linearizes the max operator internally, but it is one such way.

\end{document}